# A Density-Dependent Diffusion Model of an Interacting System of Brownian Particles

Ahmed M. Fouad [(1)], Marwa M. Fouad [(2)]

Department of Mathematics & Computer Science, La Salle University, Philadelphia, PA 19141[(1)]

Basic Sciences Department, Modern Academy for Engineering & Technology, Cairo, Egypt[(2)]

**Abstract**—Density-dependent diffusion is a widespread phenomenon in nature. We have examined the density-dependent diffusion behavior of some biological processes such as tumor growth and invasion [23]. Here, we extend our previous work by developing computational techniques to analyze the density-dependent diffusion behavior of one-dimensional interacting particle systems, which have been used to model numerous microscopic processes [17-19], and we apply our techniques to an interacting system of Brownian particles, with hard-core interactions and nearest-neighbor adhesion, known as single-file dynamics. Through large-scale numerical simulations that exploit Monte-Carlo techniques and high-performance computing resources, we show that the diffusion rate in such systems depends on the average particle density. Extensions to the techniques we present here enable researchers to examine the density-dependent diffusion behavior of many physical systems in nature that undergo one-dimensional diffusion associated with a change in the particle density; such as ion transport processes, channeling in zeolites, etc.

## 1. Introduction

Computer simulation is becoming an increasingly common resource among scientists because the accuracy of many statistical and analytical hypotheses can be tested experimentally using computer simulations. In this paper, we present computational techniques that we exploit to examine the density-dependent diffusion behavior of a one-dimensional interacting particle system known as single-file dynamics. Our computational results confirm that the diffusion in such systems is density-dependent, where the time evolutions of various motion properties, such as the diffusion coefficient, average first moment, and distribution width, are monitored and found to depend on the average particle density.

Single-file diffusion has been a major research focus for many years [1-15] after discovering that many physical systems such as ion transport through biological membranes [16,17], one-dimensional hopping conductivity [18], channeling in zeolites [19], etc., obey single-file dynamics. In single-file dynamics, Brownian particles (referred to as the tracer or tagged particles) diffuse in one-dimensional domains and collide with each other. The particles' incompressibility implies that their mutual passage is excluded; thus they maintain the same order at all times. If the average particle density is kept fixed during the diffusion, the no-passing restriction causes the famous anomalous sub-diffusion behavior [1], which originates from the fact that the motion of a tagged particle is, anywhere and at any time, hindered by collisions with all surrounding particles.

In addition to the no-passing restriction, we introduce adhesion to single-file dynamics as a second inter-particle interaction rule. In our simulations, the adhesion factor affects only neighboring particles. As we explain in detail in Sec. 3, adhesion slows the diffusion down by reducing the transition probabilities of all the particles equally; however, the anomalous sub-diffusion behavior of single-file dynamics that is observed when the average particle density is kept fixed remains almost intact [32].

Our work is outlined as follows. In Sec. 2 we give a brief introduction to the physics of diffusion, described by Fick's law, and the density-dependent diffusion phenomenon. In Sec. 3 we present an overview of our simulation model and explain how we introduce the adhesion factor to it. In Sec. 4 we discuss an analytical representation for the tracer diffusion coefficient of core particles for single-file dynamics with a Gaussian density approximation. In Sec. 5 we operate on a linear system of particles with a Gaussian density distribution and we present our computational results which include studying the trajectories of the individual tagged particles in areas with high (near the center) and low (near the tails) particle densities, the time evolution of the average first moment of various particle groups with various densities, the time evolution of the distribution width, and the dependence of the tracer diffusion coefficients of both core and tail particles on the total number of particles. Moreover, we present samples of the programming effort for our simulations including the particles' sorting algorithm [some of our programs were run on Temple Owl's Nest (a high-performance computing cluster)]. Finally, in Sec. 6 we present the summary, conclusions, and outlook.

## 2. Density-Dependent Diffusion

Diffusion is defined as the movement of a substance down a concentration gradient. The physics of diffusion is well described by Fick's law which relates the diffusive flux to the density; it postulates that the direction of the diffusive flux is from high to low density spaces, which is the opposite direction of the density gradient. In one-dimension domains, Fick's law is expressed as

$$J = -D\frac{\partial \phi}{\partial x}, \tag{1}$$

where $J$ is the diffusion flux, $D$ is the diffusion constant or diffusivity, and $\phi$ is the density. In two or more dimensions, Fick's law is expressed as

$$J = -D\nabla\phi. \tag{2}$$

A density-dependent diffusion process is a one in which the diffusion coefficient is a function of the density of the diffusing substance. Density-dependent diffusion has been observed in many physical systems; such as radial propagation in population dynamics [20], travelling wave fronts in bacterial growth models [21], periodic Lorentz gas [22], tumor growth and invasion [23], reaction-diffusion systems [24], etc.

## 3. Model and Simulation

In our model, the particles diffuse through a one-dimensional lattice with 10001 sites (cells). The single-file restriction implies that each site can be occupied by one particle at most at any moment in time. The $x$ coordinates of the leftmost, central, and rightmost sites are -5000, 0, and 5000, respectively. At time $t=0$ (initial state), we distribute the particles among the lattice sites according to a Gaussian density function of the form

$$P(x,t=0) = e^{-x^2/\sigma^2}, \tag{3}$$

where $P(x,t=0)$ is the occupancy probability of the lattice sites at $t=0$. For the central site (the site with $x=0$), $P = e^{-0/\sigma^2} = 1$ at $t=0$, independent of $\sigma$. As we approach the tails of the lattice, the occupancy probability diminishes exponentially; as a consequence, sites located near the tails are almost guaranteed to be empty at $t = 0$. The overall view of a particle system with such an occupancy probability distribution at $t = 0$ is high particle density near the center and low density near the tails. We can calculate a numerical estimate of the average number of particles in the simulation according to

$$N_{ave} = \sum_{x=x_{\min}}^{x_{\max}} e^{-x^2/\sigma^2} \ . \tag{4}$$

It is clear that $N_{ave}$ increases as $\sigma$ increases. If $1 << \sigma << x_{\max}$ (which is true in our simulations) we get

$$N_{ave} \approx \int_{-\infty}^{\infty} e^{-x^2/\sigma^2} dx = \sigma\sqrt{\pi} \ . \tag{5}$$

After the initial particle state is configured, the particle system evolves by going through a specific number of Monte-Carlo (time) steps. In each Monte-Carlo step, every particle in the system is considered once in a random order, where a hopping direction is picked at random, either to the right or left site, with a selection probability of ½ for each. The particle then hops toward the selected site; however, the transition could be successful or not depending on the two inter-particle interaction rules we consider in our model (the single-file restriction and nearest-neighbor adhesion). At a given time step, if the transition attempt fails for a specific particle, it must wait until the next time step before hopping again, where all the particles in our model have the same hopping frequency. For illustration, for a given particle, let us assume that the selected hopping direction is to the right (see Fig. 1), if the right site is already occupied, the attempt is unsuccessful, if the sites to both the right and left are vacant, the attempt succeeds with probability 1, and, finally, if the right site is vacant but the left site is occupied, the transition attempt to the right site succeeds with a reduced probability $1-\alpha$, where $\alpha$ is the adhesion coefficient, because the transition breaks a nearest-neighbor bond. Analogous rules govern hopping to the left site. The above rules can be simply expressed in terms of occupation numbers $u_j$, where $u_j = 1$ if the site $j$ is occupied and $u_j = 0$ if it is empty. The probability $T_i^+$ that a transition attempt from site $i$ to the right is successful and the corresponding probability $T_i^-$ for a transition attempt to the left are given by

$$T_i^+ = (1-u_{i+1})(1-\alpha u_{i-1})\,, \tag{6}$$

$$T_i^- = (1-u_{i-1})(1-\alpha u_{i+1})\,, \tag{7}$$

respectively.

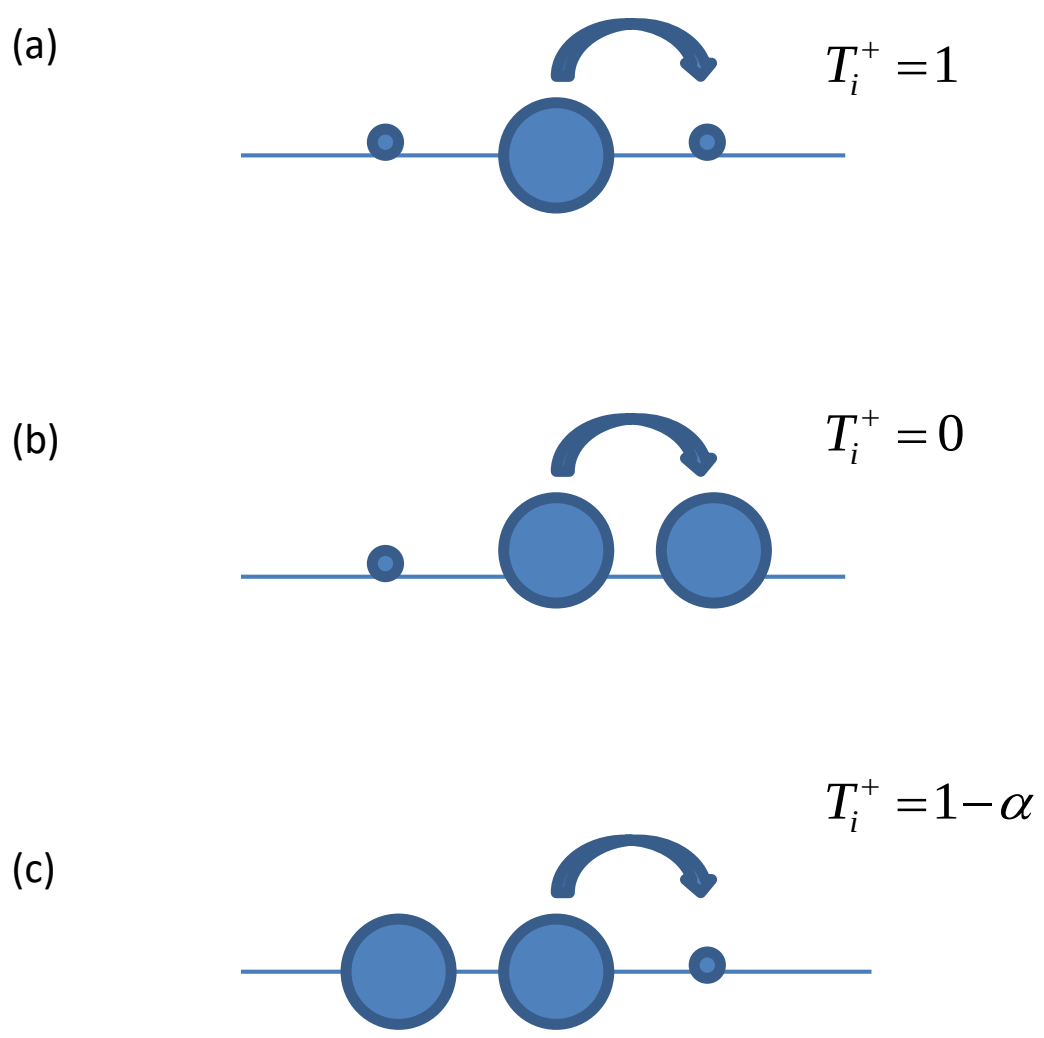

**Fig. 1**. Particle transition probabilities in single-file dynamics with two inter-particle interaction rules: the single-file restriction and nearest-neighbor adhesion. In (a) the particle hops to a vacant site; $T_i^+ = 1$. In (b) it hops to an occupied site; $T_i^+ = 0$. In (c) it hops to a vacant site; but the transition breaks a nearest-neighbor bond, so it takes place with a reduced probability $T_i^+ = 1-\alpha$.

For our numerical simulations, the average first moment, distribution width, and mean square displacement per particle are calculated according to:

$$\langle x(t)\rangle = \frac{1}{N}\sum_{i=1}^{N} x_i(t)\,, \tag{8}$$

$$W(t) = \left(\frac{1}{N}\sum_{i=1}^{N} x_i(t)^2\right)^{1/2}\,, \tag{9}$$

$$\Delta x^2 = \frac{1}{N}\sum_{i=1}^{N}(x_i(t) - x_i(0))^2 \,, \tag{10}$$

respectively, where $x_i(t)$ and $x_i(0)$ are the positions of the particles at times $t$ and 0 (initial state), respectively, and $N$ is the total number of particles.

## 4. Analytical Representation for the Tracer Diffusion Coefficient

Aslangul [25] considered the single-file case with the Gaussian density approximation, where he derived an analytical expression for the tracer diffusion coefficient $D_T$ of core particles. He considered a system of $N$ particles governed by the many-particle diffusion equation:

$$\partial_t P(X,t\,|\,X_0) = D\sum_{j=1}^{N}\partial_{x_j}^2 P(X,t\,|\,X_0)\,, \tag{11}$$

where $P(X,t\,|\,X_0)$ is the joint probability-density function for all the particles in the file and $D$ is the diffusion constant of the entire system (we have shown analytically that $D = 0.5$ for single-file diffusion with the symmetric simple exclusion process [23], which is the case we consider in our simulations). Aslangul analyzed the case in which the particles are all initially concentrated at the origin. For this initial condition, the solution to the diffusion equation (11) is

$$P(x_1, x_2, ..., x_N; t) = N!\prod_{n=1}^{N}\frac{e^{-x_i^2/(4Dt)}}{\sqrt{4\pi Dt}}\,, \tag{12}$$

for $x_1 < x_2 < ... < x_N$, and $P(x_1, x_2, ..., x_N; t) = 0$ otherwise. This follows from solving Eq. (11) with all the initial positions $x_{0,1}, ..., x_{0,N}$ set equal to zero. From Eq. (12) Aslangul derived an exact expression for the one-particle probability density of particle $n$ as

$$P_n^1(x,t) = \int_{-\infty}^{\infty} dx_1 \int_{x_1}^{\infty} dx_2 \int_{x_2}^{\infty} dx_3 ... \int_{x_{N-1}}^{\infty} dx_N\, \delta(x - x_n) P(x_1, x_2, ..., x_N; t)\,. \tag{13}$$

With a Gaussian approximation for the density, for the middle particle in a file of $N$ particles (where $N$ is odd), he obtained

$$\Delta x_{(N+1)/2}^2 \approx \frac{\pi}{N} Dt\,. \tag{14}$$

From the famous Einstein relationship, the mean square displacement of a Brownian particle in one-dimension can be expressed as

$$\Delta x^2 = 2D_T t \,. \tag{15}$$

Comparing Eqs. (14) and (15), we see that $D_T$ of the middle particle is proportional to 1/N.

## 5. The Computational Results

Since the emergence of scientific computing, the sorting problem has been a research focus; because, although apparently simple, it is quite complicated to be solved analytically. A sorting algorithm is an algorithm that puts the elements of a list in a certain order. The time complexity of a sorting algorithm is a measure of the amount of time taken by the algorithm to run as a function of the input length. The input passed to the sorting algorithm is a string of characters. In our model, the string input is the number of particles $N$ to undergo single-file diffusion in a lattice. According to Eq. (5), if $\sigma = 250$, we get approximately 443 particles (3 characters as the input string length). Consider a list that has $N$ elements, for many sorting algorithms such as Quicksort [26], Merge sort [27], Tournament sort [28], Heapsort [29], Introsort [30], Binary tree sort [31], etc., $N \log N$ is found to be the best time complexity.

We have created a C function (see Fig. 2) that creates a new, maximally scrambled list of particle indices each time step in the simulation. For a particular step, the order of the generated list elements (from left to right) is the order at which the particles hop at that step.

```
void
set_indx(int *pindx, int nparts)
{

int i, *nvisited, nadd=0;

nvisited=(int*)calloc(nparts, sizeof(int));

for (i= 0; i<nparts; i++) nvisited[i]=0;

    for(;;)
    {
        i=drand48()*nparts;
        if(nvisited[i]!=0)continue;
        nvisited[i]++;
        pindx[nadd]=i;
        nadd++;
        if(nadd==nparts)break;
    }
free((void*)nvisited);
return;
}
```

**Fig. 2**. C code of a function that creates a maximally-scrambled set of particle indices.

We make the following remarks on the function “set_indx” presented in Fig. 2:

1- The function “set-indx” creates a maximally-scrambled list of particle indices.

2- Here, “*nvisited” is an array (created by a pointer) that has $N$ elements. Each element can be either 0 or 1, which represents the number of times a particle hops at a time step.

3- The allocated memory for “*nvisited” is freed at the end of the function using the statement “free((void*)nvisited);”.

4- For each iteration in the “for(;;)” loop, a random-number generator “drand48()” function is used to draw an integer number “i” between 0 and $N-1$.

5- Note that “nvisited[i]=0” if this is the first time the number “i” is generated, “nvisited[i]=1” if it is the second time, and “nadd” represents the current number of successful, first-time random numbers generated.

6- If “nvisited[i]=0”, the particle index is stored in the “pindx” array at the order it was generated according to “pindx[nadd]=i”, then “nadd” is incremented by the statement “nadd++;” before shifting to the next iteration in the loop.

7- If “nvisited[i]=1”, this means that the random number has been generated before within the same time step and, as a consequence, it is disqualified because each particle can hop only once in any time step, and the loop iterates again to generate a new number.

8- The loop stops iterating once the total number of the random numbers generated (between 0 and $N-1$) becomes equal to the total number of particles in the system $N$; see the statement “if(nadd==nparts)break;” in the program.

9- Suppose we have a distribution of 10 particles. The generated list by the code in Fig. 2 might, for example, look like {8,6,3,2,4,5,9,1,7,0}, which implies that for this particular time step, the particle with index 8 hops first and the particle with index 0 (the leftmost particle in the distribution) hops last.

10- The function “set_indx” is called for each time step in the simulation and a new corresponding list of particle indices is generated.

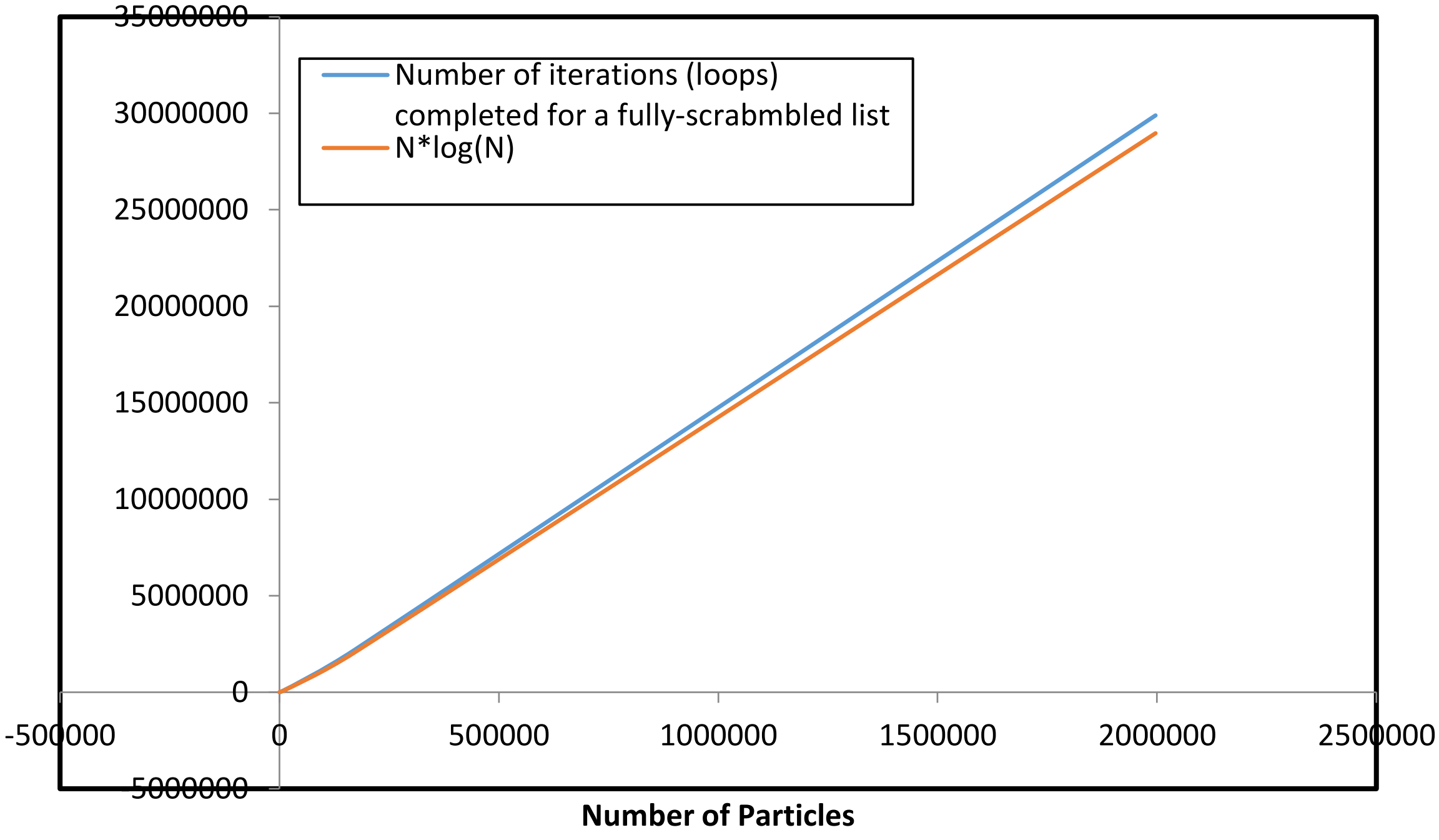

**Fig. 3.** Comparison of the time complexities. The top blue line represents the time complexity of our C function in Fig. 2. The bottom red line represents the $N \log N$ time complexity.

In Fig. 3 we compare the time complexity of our C function in Fig. 2 to the $N \log N$ time complexity. The top blue curve represents the number of times the "for(;;)" loop in the C function has to iterate in order to generate a maximally-scrambled set of particle indices as a function of the total number of particles. The bottom red line represents the $N \log N$ time complexity. It is obvious that the two curves are almost identical. This shows that our function has the same best $N \log N$ time complexity for maximum scrambling as for the sorting algorithms referred to earlier.

By studying the time evolution of $\langle x(t) \rangle$, calculated according to Eq. (8), for the Gaussian distribution defined in Eq. (3), we show in Fig. 4 that the location of the center of mass of particle systems undergoing single-file diffusion remains almost invariant with time; because at each time step, every particle is equally likely to attempt a transition to the right as to the left site and, moreover, it vanishes if the distribution is initially centered on the origin (which is the case in our simulations).

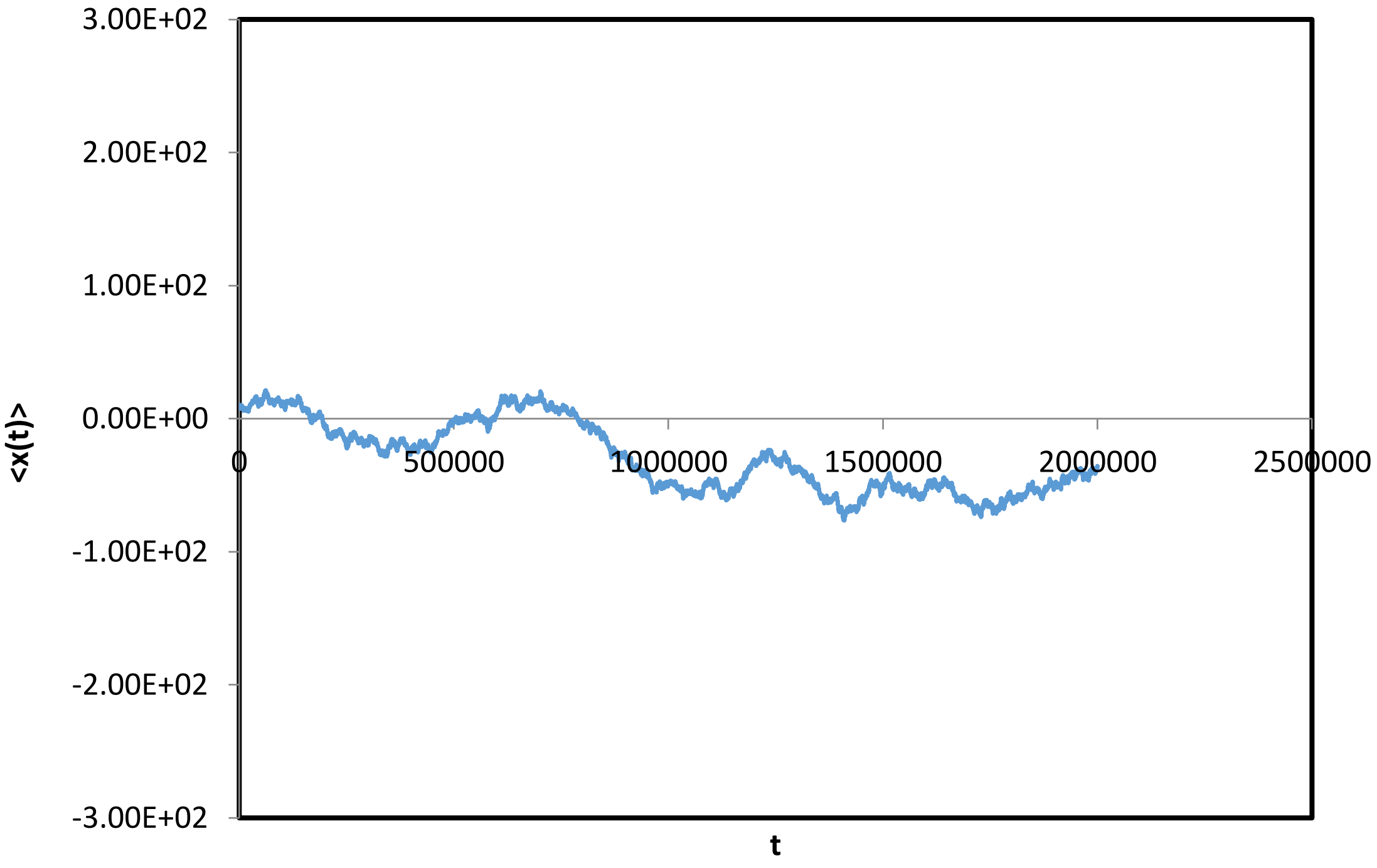

**Fig. 4.** The location of the center of mass in single-file diffusion is almost time-invariant; that is, the diffusion is symmetric around the center of mass. Here, we operate on a Gaussian density distribution with $\sigma = 250$ ( $N = 449$) and $\alpha = 0.1$.

We see in Fig. 5 that a reduction in the average particle density takes place during the diffusion, where the single-file restriction directs the diffusion of the individual tagged particles away from the crowded center and toward the empty tails of the Gaussian distribution, resulting in a noticeable divergence in the particle trajectories. The question we are trying to answer is whether the diffusion is density-dependent or not. Since $\langle x(t)\rangle$ of the entire distribution vanishes, the technique we use to investigate the density-dependent diffusion behavior is to pick a particle group, either on the right or left side of the origin, then monitor the time evolution of $\langle x(t)\rangle$ for that particular group. The slope of the curve of $\langle x(t)\rangle$ vs. $t$ is indicative of the average propagation speed at which that group is diffusing through the lattice toward either end. If $d\langle x(t)\rangle / dt$ is not constant; that is, decreases with time as the average particle density decreases due to diffusion in the lattice, then the diffusion is density-dependent.

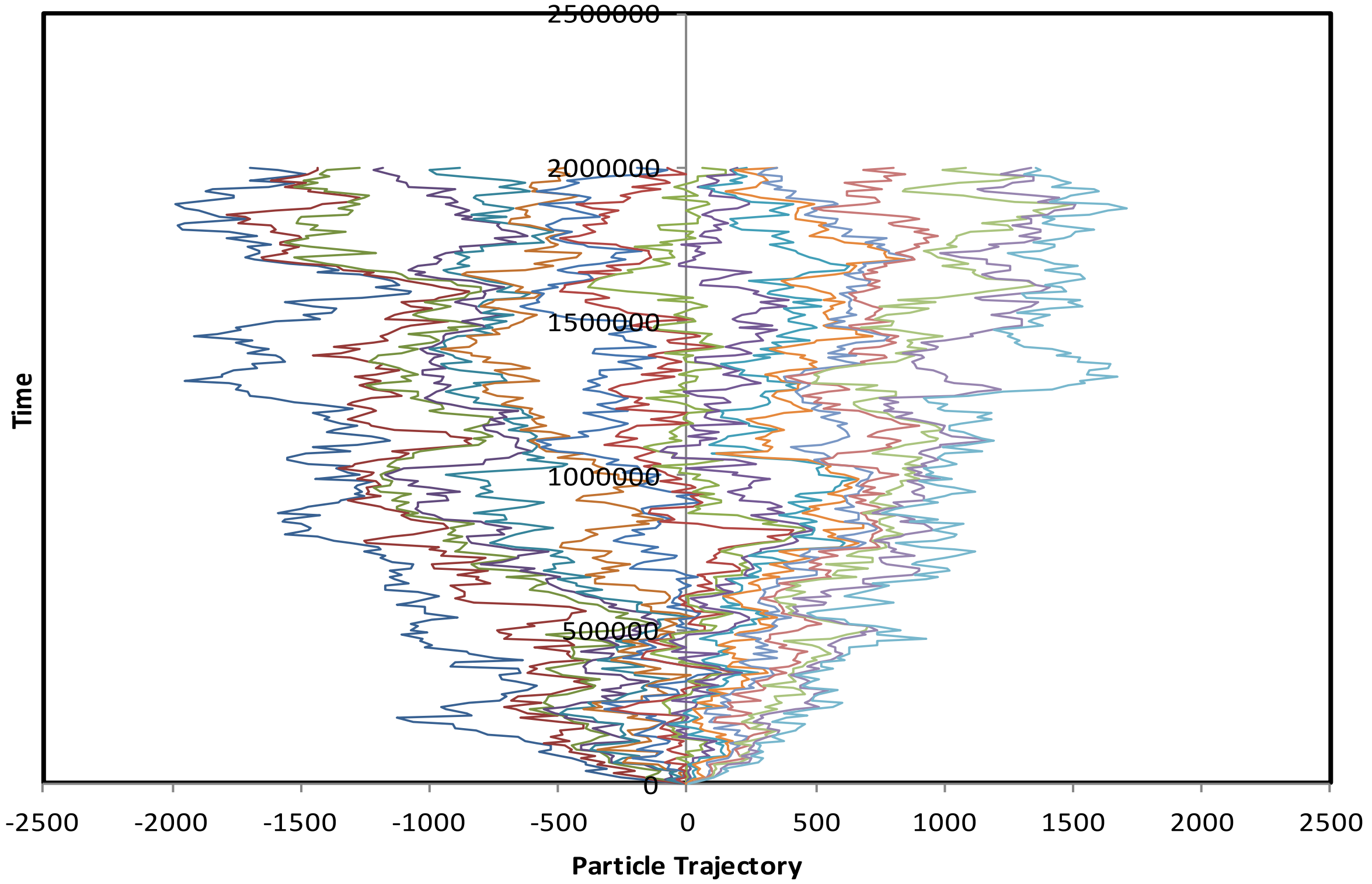

**Fig. 5.** Time evolution of the particle trajectories. The single-file restriction directs the diffusion of the individual tagged particles away from the crowded center of the Gaussian distribution towards the empty tails resulting in a progressive drop in the average particle density with time.

In Fig. 6 we operate on a Gaussian distribution with $\sigma \approx 282.1$ ($N = 500$) and $\alpha = 0.1$. We divide the 500 particles into ten groups of 50 particles each. The first particle group has the 50 leftmost particles in the distribution and the tenth particle group has the 50 rightmost particles. The indices of the leftmost and rightmost particles are $P = 0$ and $P = 499$, respectively. We then pick the 50-particles group with indices $P \in \{400, 449\}$ (the ninth particle group) and monitor the time evolution of $\langle x(t) \rangle$ for that group, where in Fig. 6(a) we show the time evolution of $\langle x(t) \rangle$ for 20 different simulations and in Fig. 6(b) we show the average of the 20 simulations in 6(a). We see that $d\langle x(t) \rangle / dt$ decreases progressively with time as the average particle density decreases, confirming the density-dependent diffusion behavior. The ninth particle group is close to the right tail of the distribution and is considered to be in a medium-to-low density region.

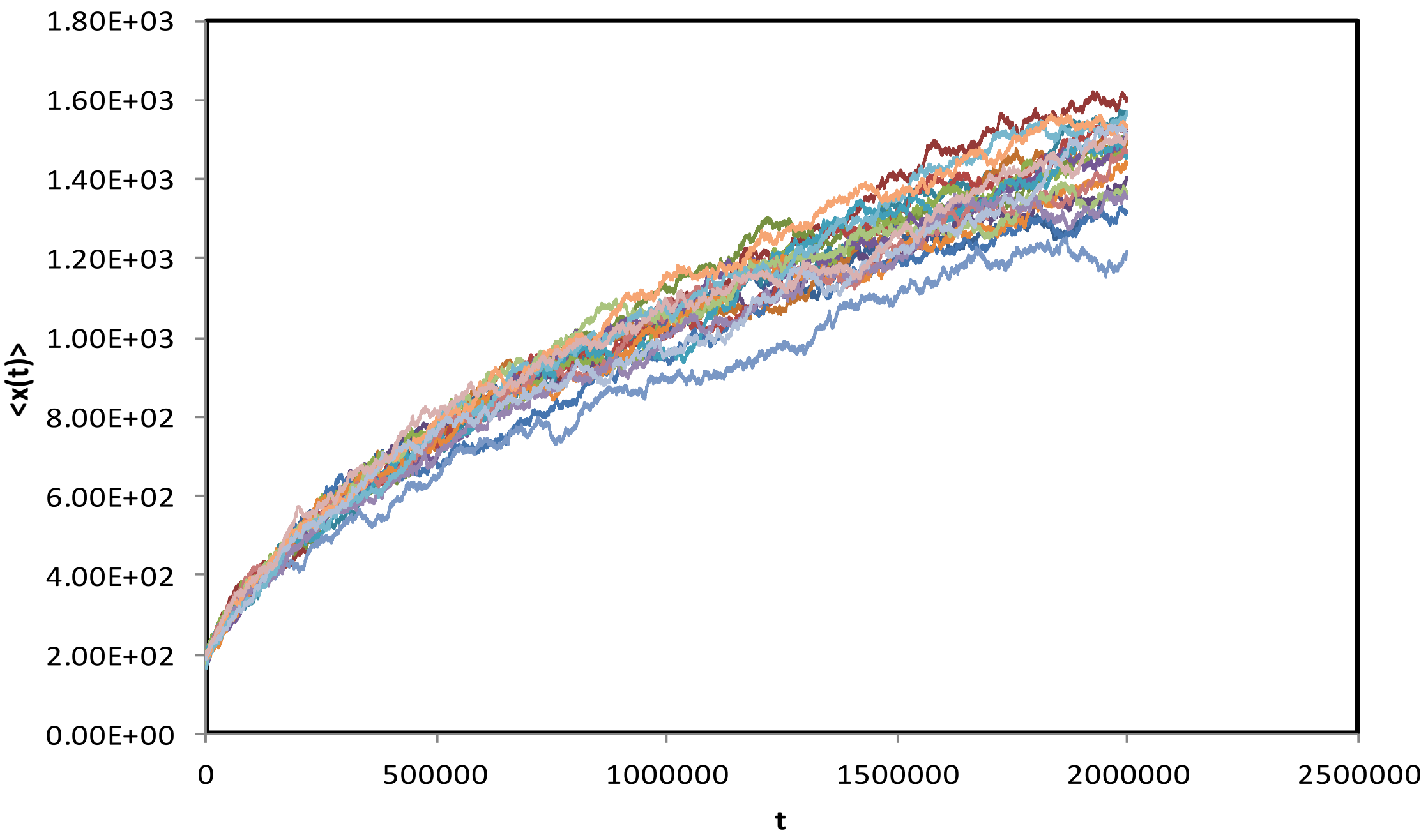

(a)

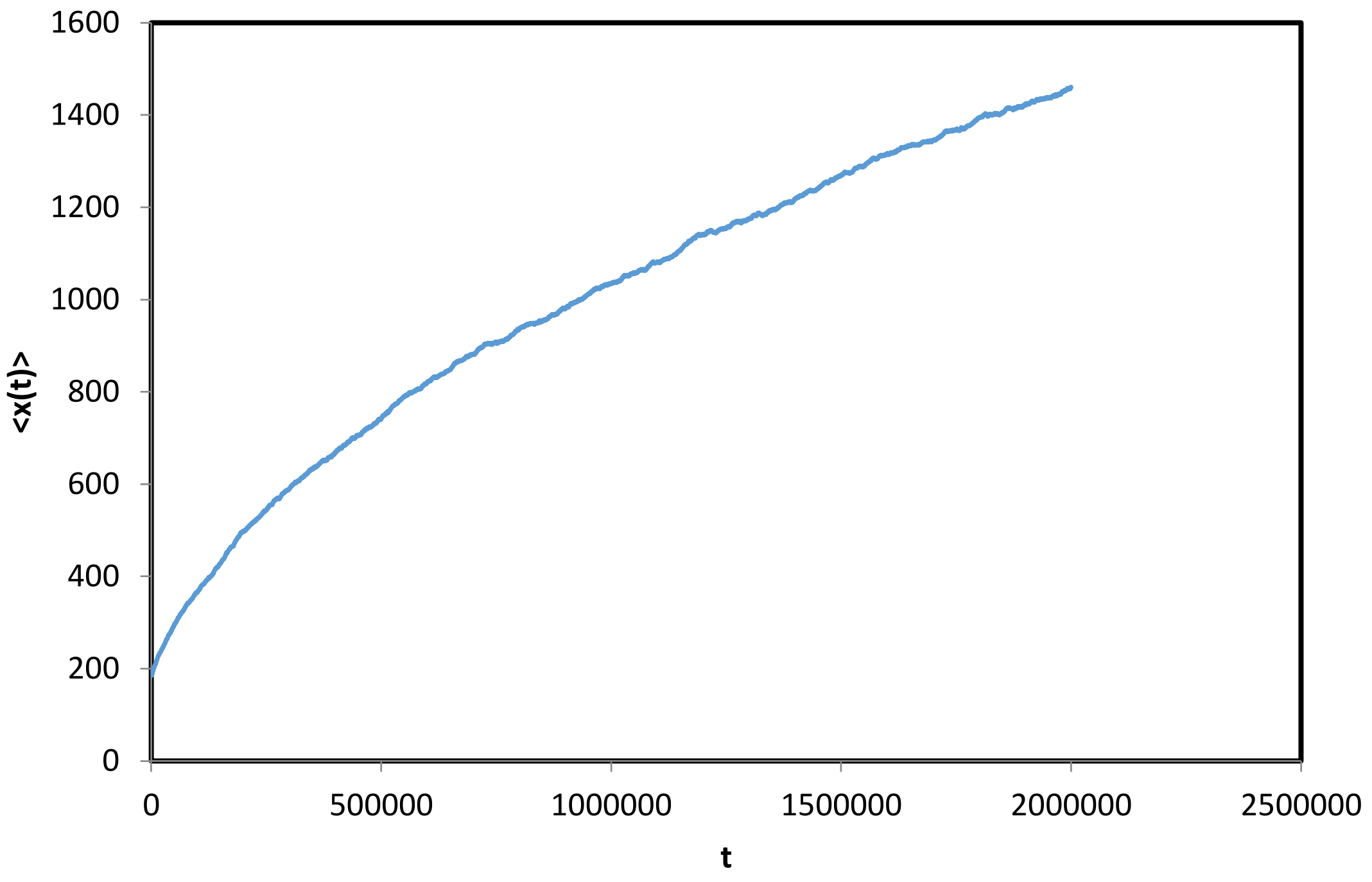

(b)

**Fig. 6.** Time evolution of the average first moment for a 50-particles group located on the right side of the center of the Gaussian distribution. In (a) we present our computational results for 20 different simulations. In (b) we show the average curve of the 20 simulations in (a), where we can notice that the slope $d\langle x(t)\rangle / dt$, which is indicative of the average propagation speed of this group toward the right end of the lattice, is progressively decreasing with time, as the average particle density decreases due to the diffusion, confirming a density-dependent diffusion behavior.

Because of the single-file restriction, particles on the right edge of the Gaussian distribution tend to diffuse to the right and particles on the left edge tend to diffuse to the left. We expect that when the average particle density decreases as the particles get further away from one another, the impact of the single-file restriction on the diffusion subsequently decreases and, as a consequence, the diffusion rate of the entire particle distribution toward the two ends of the lattice decreases. Moreover, the diffusion of the individual tagged particles is faster at the tails of the distribution relative to the core due to a greater availability of space available for the diffusion.

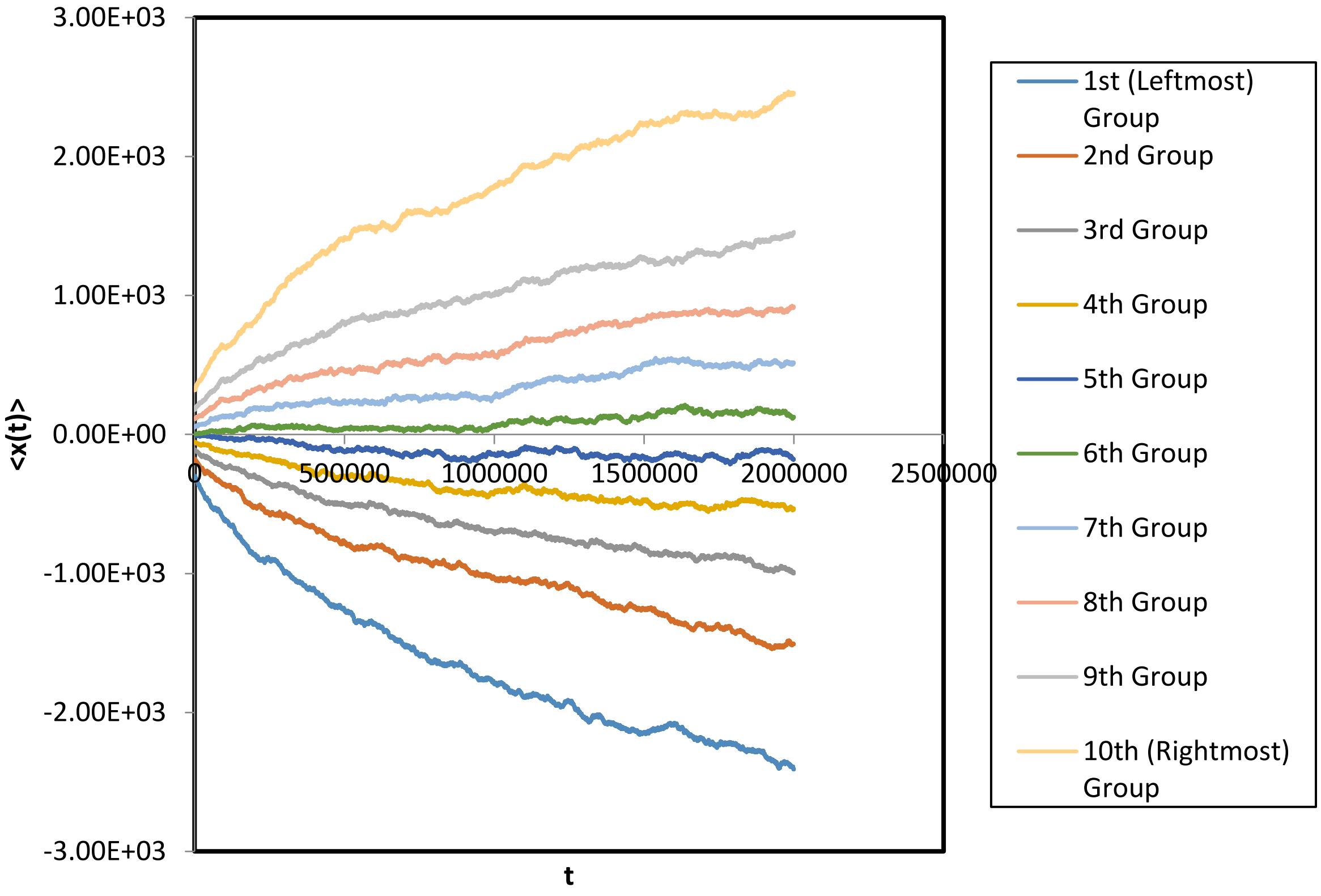

**Fig. 7.** Time evolutions of the average first moment for various particle groups in the Gaussian distribution. A density-dependent diffusion behavior is confirmed for all ten particle groups, where the magnitude of $d\langle x(t)\rangle / dt$ decreases with time for all ten curves, as the corresponding average particle densities decrease.

In Fig. 7 we operate on the same Gaussian distribution as in Fig. 6, and we show here the time evolutions of $\langle x(t)\rangle$ for the ten 50-particles groups mentioned earlier. The top half of Fig. 7 depicts the time evolutions of $\langle x(t)\rangle$ for the five particle groups located to the right of the center of the distribution and the bottom half depicts the corresponding time evolutions for the five particle groups located to the left. We see that the density-dependent diffusion behavior is preserved in all regions of the distribution, where the magnitude of $d\langle x(t)\rangle / dt$ decreases with time for all ten curves. Moreover, the diffusion is faster at the tails relative to the core due to a greater abundance of free space available for the diffusion, where the two terminal (first and tenth) particle groups have the greatest magnitudes of $d\langle x(t)\rangle / dt$, and the two innermost (fifth and sixth) groups have the smallest magnitudes at any moment in time.

```
for (p=0;p<50;p++) AverageFirstMoment1+=xcell[cell[p]]/50;
for (p=50;p<100;p++) AverageFirstMoment2+=xcell[cell[p]]/50;
for (p=100;p<150;p++) AverageFirstMoment3+=xcell[cell[p]]/50;
for (p=150;p<200;p++) AverageFirstMoment4+=xcell[cell[p]]/50;
for (p=200;p<250;p++) AverageFirstMoment5+=xcell[cell[p]]/50;
for (p=250;p<300;p++) AverageFirstMoment6+=xcell[cell[p]]/50;
for (p=300;p<350;p++) AverageFirstMoment7+=xcell[cell[p]]/50;
for (p=350;p<400;p++) AverageFirstMoment8+=xcell[cell[p]]/50;
for (p=400;p<450;p++) AverageFirstMoment9+=xcell[cell[p]]/50;
for (p=450;p<500;p++) AverageFirstMoment10+=xcell[cell[p]]/50;
```

**Fig. 8.** C code for calculating the time evolutions of the average first moments for the ten particle groups in Fig. 7.

In Fig. 8 we present the C code we developed and used to calculate the time evolutions of $\langle x(t)\rangle$ for the ten particle groups in Fig. 7. We make the following remarks:

1- Ten double-precision floating-point variables are declared, a variable for each particle group. These are named “AverageFirstMoment1”, “AverageFirstMoment2”,…, and “AverageFirstMoment10”, in which the values of the particle groups’ moments are stored.

2- Before the beginning of each time step, the values of the ten variables are set equal to zero to avoid numerical accumulation.

3- In each time step, the sum of the $x$ coordinates, “xcell[cell[p]]”, of the particles of each group is stored in the corresponding variable, then divided by 50 for averaging, where each group has 50 particles.

4- Note that “cell[p]” is the sites (cells) indices which can be anywhere between 0 and 10000, where we operate on a lattice with 10001 sites, as mentioned in Sec. 3 in detail.

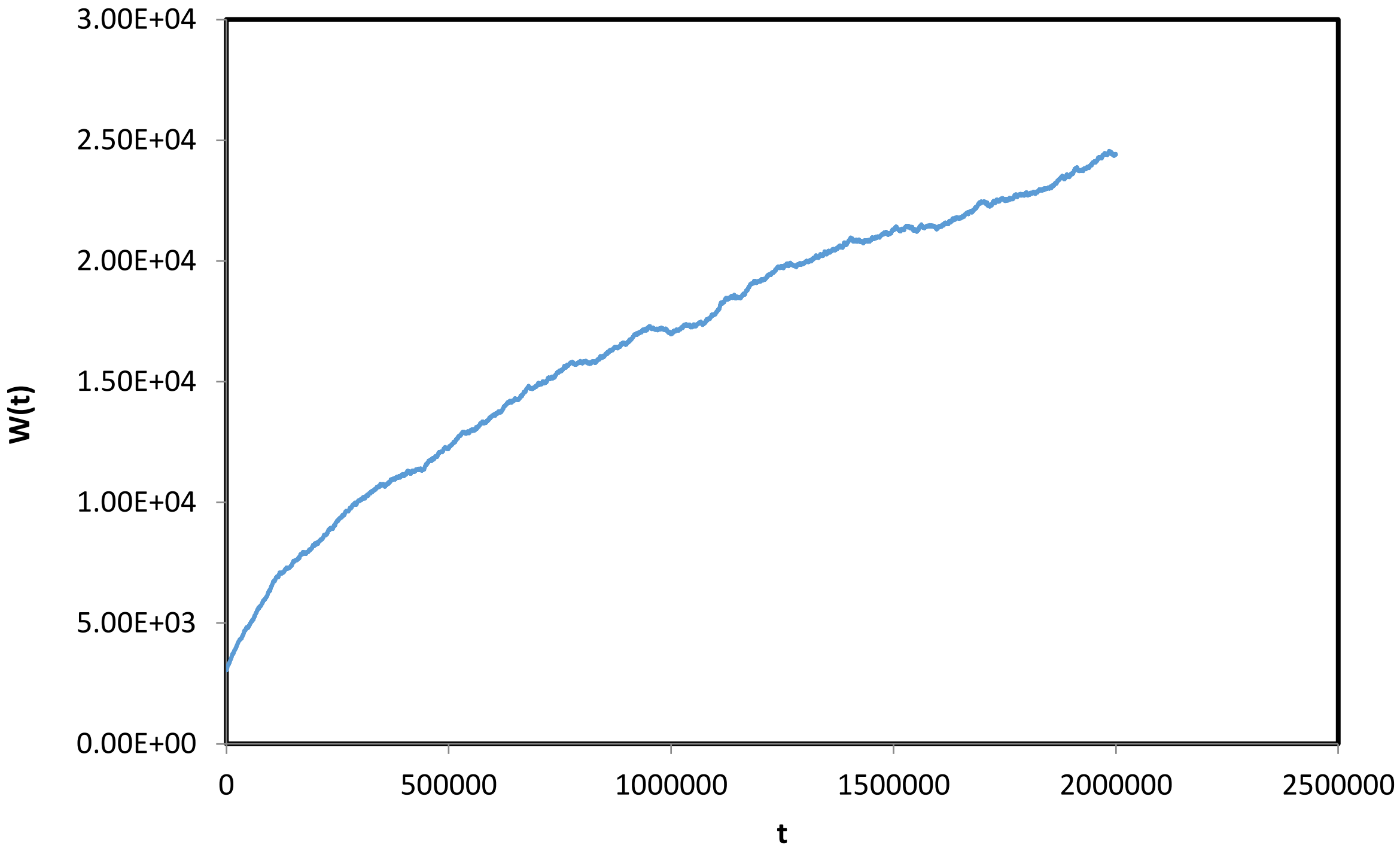

**Fig. 9.** Examining the time evolution of the distribution width confirms the density-dependent diffusion behavior, where $dW(t)/dt$, which is indicative of the diffusion rate through the lattice, decreases with time as the particles approach the two ends of the lattice.

The density–dependent diffusion behavior is confirmed once again in Fig. 9 by examining the time evolution of the distribution width $W(t)$, calculated at each time step according to Eq. (9), where we operate on a Gaussian distribution with $\sigma = 250$ ($N_{parts} = 462$) and $\alpha = 0.1$, and we see that the slope of the curve $dW(t)/dt$, which is indicative of the diffusion rate through the lattice, decreases with time, as the average particle density decrease, confirming once again a density-dependent diffusion behavior.

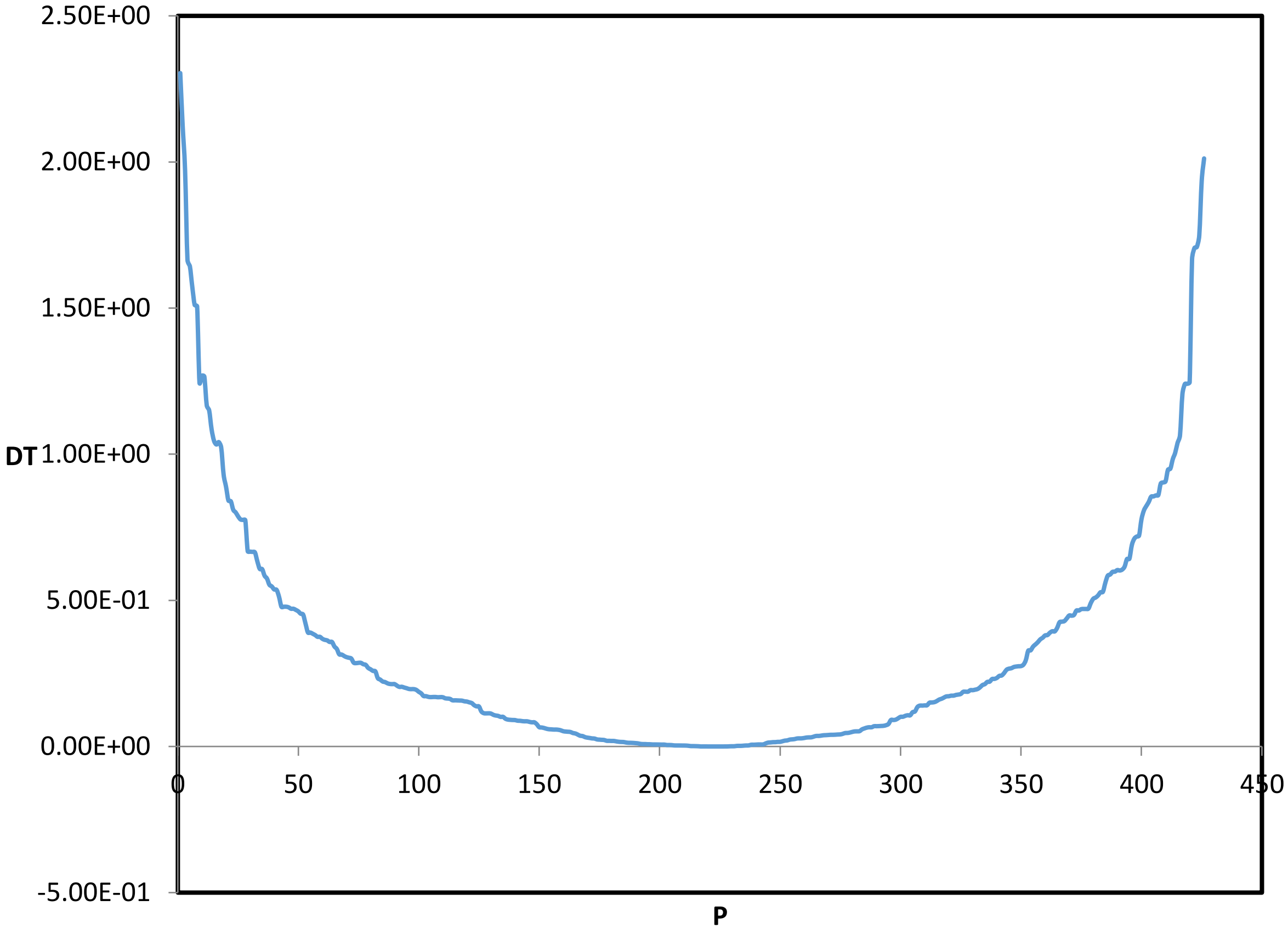

**Fig. 10.** Measurement of the tracer diffusion coefficient values at late times for all the particles in the Gaussian distribution. We see that tail particles have much greater $D_T$ values relative to their core counterparts.

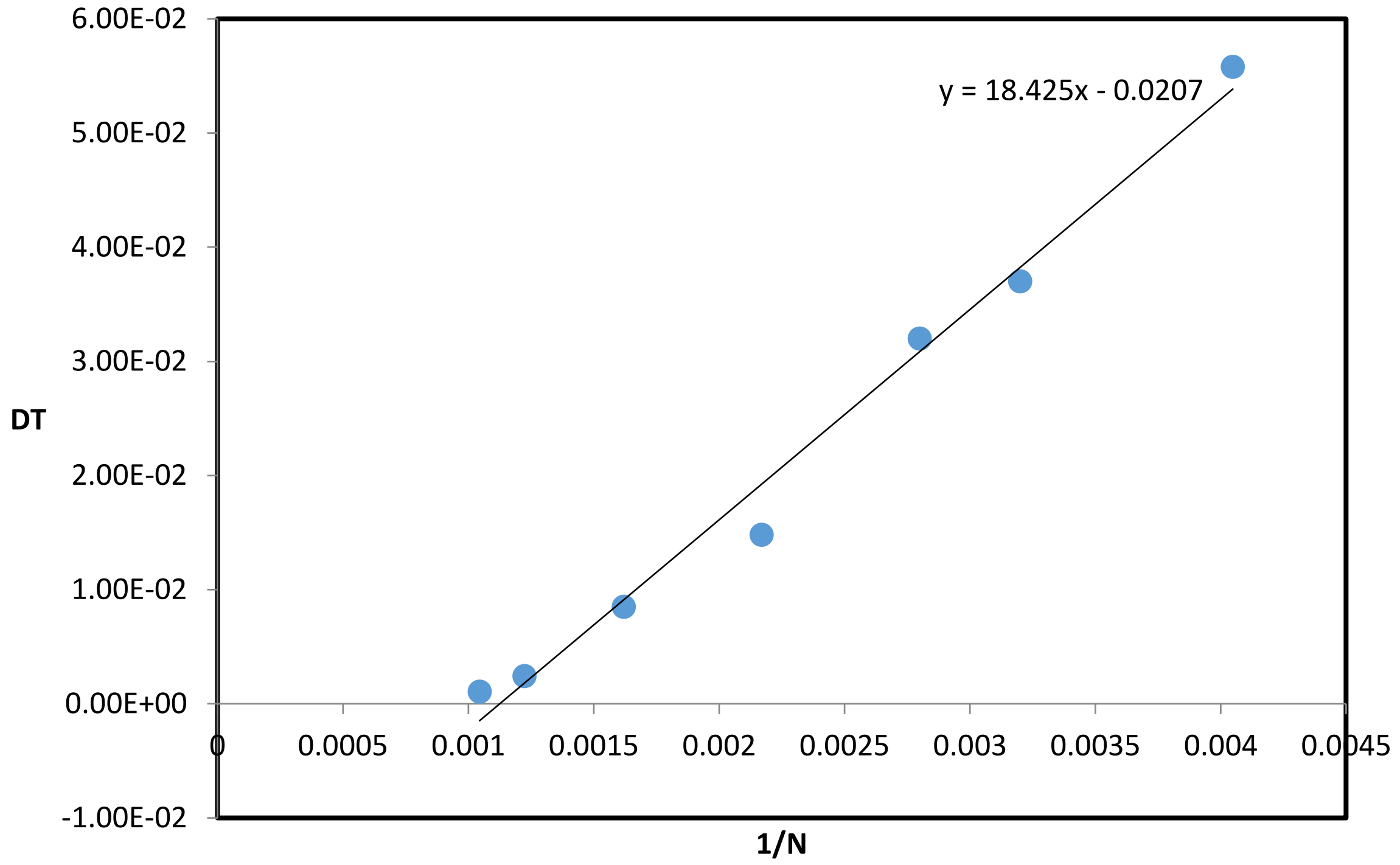

(a)

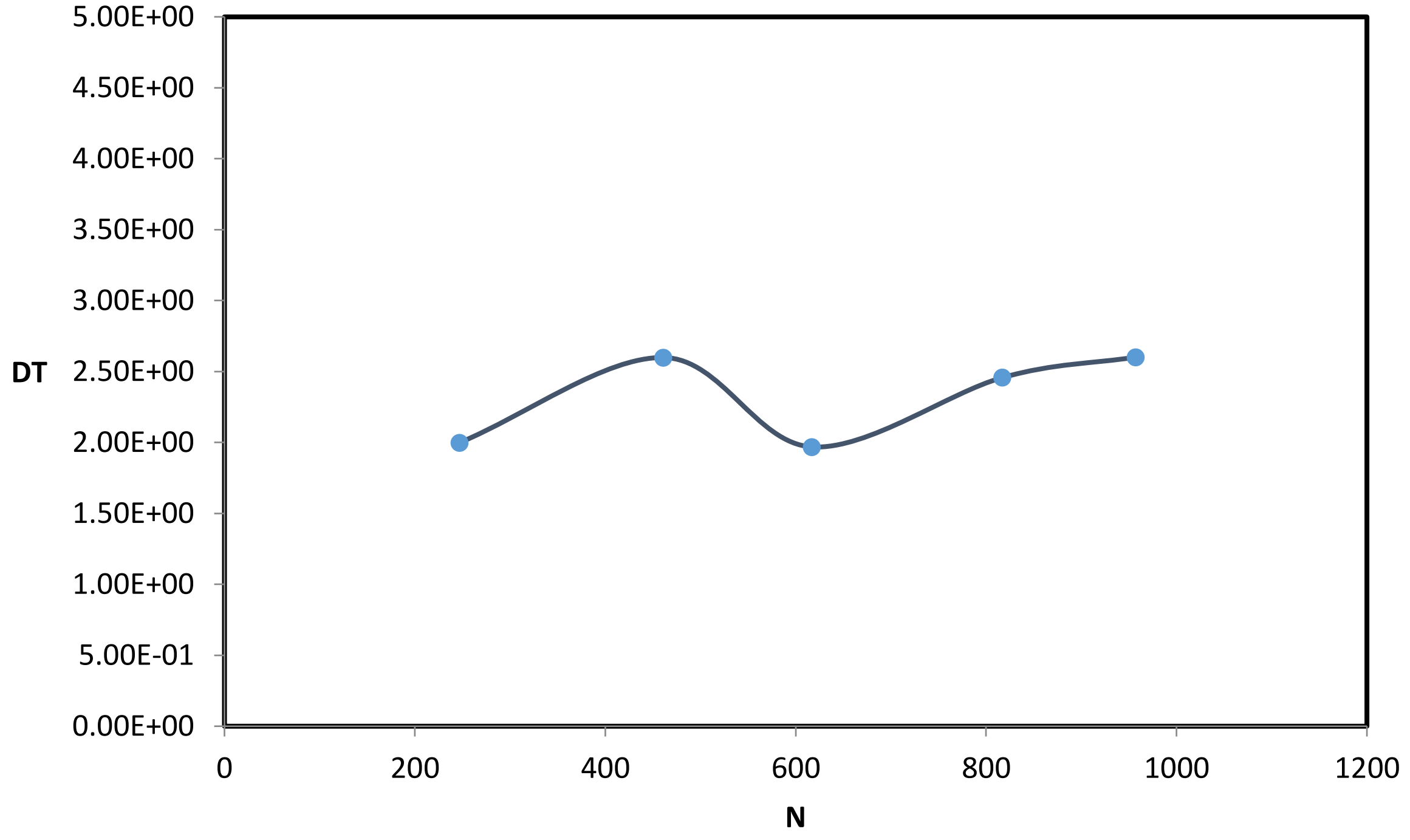

(b)

**Fig. 11.** (a) Variation of the tracer diffusion coefficient of core particles with $1/N$. (b) Variation of the tracer diffusion coefficient of tail particles with $N$. The data in (a) and (b) are based on averaging over 20 simulations.

In Fig. 10 we operate on a Gaussian distribution with $\sigma = 240$ ($N = 426$), and we measure the values of $D_T$, calculated according to Eq. (15), for all 426 particles at late times, where we plot $D_T$ vs. the particle indices $P$ ($P = 0$ represents the leftmost particle and $P = 425$ represents the rightmost one). We see that tail particles have the greatest $D_T$ values (relative to their core counterparts), indicating that the diffusion of the individual tagged particles is faster at the tails relative to the core due to a greater availability of free space available for the diffusion. Our numerical results in Fig. 11(a) confirm the analytical prediction in Eq. (14), where we show that, for core particles, $D_T$ is directly proportional to $1/N$; however, as depicted in Fig. 11(b), for tail particles, $D_T$ does not seem to have any dependence on $N$; because tail particles don't experience the impact of the single-file restriction and inter-particle adhesion as much as their core counterparts do; that is, increasing $N$ means a more retarded diffusion for only core particles.

## 6. Summary, Conclusions, and Outlook

We have provided in this article a detailed computational methodology that can be used to analyze the density-dependent diffusion behavior of systems that undergo one-dimensional diffusion associated with a change in the average particle densities. The system we have considered here is a one-dimensional interacting particle system that undergoes single-file diffusion. In addition to the no-mutual-passage exclusion, we have introduced nearest-neighbor adhesion [32,33] as a second inter-particle interaction rule (see Eqs. (6) and (7) and Fig. 1). We have shown here by considering these two inter-particle interaction rules that for an initial Gaussian density distribution of particles [see Eq. (3)], the single-file restriction directs the diffusion of the individual tagged particles away from the distribution center (area of maximum density) toward the empty tails, resulting in a progressive drop in the average particle density with time (see Fig. 5). Studying the time evolutions of both the average first moments of the particles (see Figs. 6 and 7) and distribution width (see Fig. 9) conveys a progressive reduction in the diffusion rates of these particle systems as their average densities decrease, confirming a density-dependent diffusion behavior. Moreover, we have shown that tail particles have greater diffusion coefficients than their core counterparts (see Fig. 10). For core particles, studying the variation of $D_T$ with $1/N$ confirms a linear relationship [see Fig. 11(a)], as analytically predicted by Eq. (14); however, for tail particles, $D_T$ does not seem to have any dependence on $N$ [see Fig. 11(b)].

We have also shown that the diffusion in single-file dynamical systems is symmetric around the initial center of mass (see Fig. 4), where the time evolution of the average first moment of an entire particle distribution undergoing single-file diffusion remains almost invariant, and even vanishes if the distribution is initially symmetrically distributed around the origin. This is physically interpreted by the fact that at any moment in time, every particle is equally likely to attempt a transition to either the right or left location.

The code we have developed in Fig. 2 can be used as a computational tool for maximally-scrambled and time-efficient sorting of list elements. We have shown in Fig. 3 that our code has the best time complexity $N \log N$ as other sorting algorithms; such as Quicksort [26], Merge sort [27], Tournament sort [28], Heapsort [29], Introsort [30], and Binary tree sort [31].

As for future work, it would be informative to develop computational techniques to examine the density-dependent diffusion behavior of interacting particle systems in two and three dimensions. Moreover, to consider a more sophisticated model of inter-particle interactions; such as a one in which the adhesion is a function of the separation distance between the particles.

## Acknowledgments

We thank Dr. Theodore Burkhardt and Dr. Edward Gawlinski for very thoughtful comments and discussions.